\newcommand{\PaperTitle}{immUNITY: Detecting and Mitigating Low Volume \& Slow Attacks with Programmable Switches and SmartNICs}

\documentclass[10pt,sigconf,letterpaper,nonacm]{acmart}

\usepackage{xspace}
\usepackage{amsmath,amsfonts}
\usepackage{graphicx}
\usepackage{grffile}
\usepackage[dvipsnames]{xcolor}
\usepackage{xspace}
\usepackage{url}
\usepackage{booktabs}
\usepackage{caption}
\usepackage{subcaption}
\usepackage{titlesec}
\titlespacing{\section}{0pt}{*0.75}{*0.5}  
\titlespacing{\subsection}{0pt}{*0.5}{*0.4}  
\titlespacing{\subsubsection}{0pt}{*0.5}{1ex}  

\usepackage{enumitem}
\setlist{topsep=2pt, partopsep=0pt, itemsep=1pt, parsep=0pt}

\usepackage{algorithm}
\usepackage[noend]{algpseudocode}

\newcommand{\name}{immUNITY\xspace}
\newcommand{\off}{OFF\xspace}
\newcommand{\snic}{SmartNIC\xspace}
\newcommand{\snics}{SmartNICs\xspace}
\newcommand{\head}[1]{\vspace{1mm}\noindent {\bf #1}}
\newcommand{\subhead}[1]{\vspace{1mm}\noindent {\em #1}}

\newcommand{\eg}{\textit{e.g.,}\xspace}

\begin{document}

\title{\PaperTitle}

\author{Cuidi Wei}
\affiliation{
  \institution{The George Washington University}
  \city{Washington}
  \state{D.C.}
  \country{USA}
}
\email{cuidi@gwu.edu}

\author{Shaoyu Tu}
\affiliation{
  \institution{UC Riverside}
  \city{Riverside}
  \state{CA}
  \country{USA}
}
\email{stu024@ucr.edu}

\author{Daiki Hata}
\affiliation{
  \institution{Osaka University}
  \city{Suita}
  \state{Osaka}
  \country{Japan}
}
\email{d-hata@ist.osaka-u.ac.jp}

\author{Toru Hasegawa}
\affiliation{
  \institution{Shimane University}
  \city{Matsue}
  \state{Shimane}
  \country{Japan}
}
\email{t_hasegawa@mat.shimane-u.ac.jp}

\author{Yuki Koizumi}
\affiliation{
  \institution{Osaka University}
  \city{Suita}
  \state{Osaka}
  \country{Japan}
}
\email{ykoizumi@ist.osaka-u.ac.jp}

\author{K. K. Ramakrishnan}
\affiliation{
  \institution{UC Riverside}
  \city{Riverside}
  \state{CA}
  \country{USA}
}
\email{kkrama@ucr.edu}

\author{Junji Takemasa}
\affiliation{
  \institution{Osaka University}
  \city{Suita}
  \state{Osaka}
  \country{Japan}
}
\email{j-takemasa@ist.osaka-u.ac.jp}

\author{Timothy Wood}
\affiliation{
  \institution{The George Washington University}
  \city{Washington}
  \state{D.C.}
  \country{USA}
}
\email{timwood@gwu.edu}

\begin{abstract} 
Our analysis of recent Internet traces shows that up to 71\% of flows contain suspicious behaviors indicative of low-volume network attacks such as port scans. However, distinguishing anomalous traffic in real time is challenging as each attack flow may comprise only a few packets. 

We extend prior work that tracks heavy hitter flows to also detect low-volume and slow attacks by combining the capabilities of both switches and SmartNICs.
We flip the usual design approach by proposing an efficient filter data structure used to quickly route traffic marked as \emph{benign} towards destination end-systems. We make careful use of limited programmable switch memory and pipeline stages, and complement them with SmartNIC resources to analyze the remaining traffic that may be anomalous.
Using machine learning classifiers and intrusion detection rules deployed on the SmartNIC, we identify malicious source IPs, which then  undergo more detailed forensics for attack mitigation. Finally, we develop a dataplane based protocol to rapidly coordinate data structure updates between these devices.
We implement immUNITY in a testbed with Tofino v1 switch and Bluefield 3 SmartNIC, demonstrating its high accuracy, while minimizing traffic that's analyzed outside the switch.\footnote{We will be releasing immUNITY as open-source, and will publicly release the labeled traces in the near future}
\end{abstract}

\maketitle
\section{Introduction}
\label{sec:intro}

Designing a cost-efficient and accurate traffic monitoring infrastructure is crucial to ensure the integrity of hosted services. 
Going beyond crude volumetric attacks that overwhelm the network through sustained network activity (e.g.,  denial of service), our work seeks to identify more sophisticated {\em low volume} attacks that probe for system weaknesses (\eg SSH Brute-forcing~\cite{ssh_bruteforce}, port scan~\cite{PortScan}) or {\em slow} attacks that exploit protocol dynamics (\eg low-rate TCP or HTTP attacks~\cite{10.1145/863955.863966}).
Our analysis of Internet traces from MAWI~\cite{mawi-traffic-archive} and CAIDA~\cite{CAIDA_2019_trace} reveal that while each of these attack flows may be very low volume, they collectively represent a large fraction of total traffic---up to 71\% of flows and at least 16\% of packets. 
While such a high fraction of traffic being low volume attacks may come as a surprise, it is supported by historical observations and recent trace analysis.

Allman et al.~\cite{brief_history_scanningimc07} analyzed network traffic collected at Lawerence Berkeley National Laboratory from 1994 to 2006 and found a fundamental shift from most connections being legitimate to most being part of scanning activity, especially in 2001 when the Code Red and Nimda worm outbreaks occurred. These new automated and self-replicating attacks led to increasing scanning rates, with over two thirds of connection attempts in Dec. 2006 being scans originating from just over 1\% of the remote hosts contacting LBNL. This trend has continued, with Mazel~\cite{profiling_internet_scanners} observing continuous scanning of popular services such as SSH , HTTP, and SMB, and special ports or services linked to known vulnerabilities, throughout the MAWI~\cite{mawi-traffic-archive} trace data set from 2001 to 2016. These studies show that scanning traffic is a long lasting and pervasive problem in the Internet.

Scanning attacks have evolved with new tools, attack types, and targets. 
ZMap~\cite{zmap} and Masscan~\cite{masscan} have  fundamentally changed the ease of scanning the Internet, allowing wide ranging scans to be performed in minutes from a single IP address~\cite{demystifying_service_discovery_imc10}. 
The  Internet-of-Things (IoT) botnet Mirai~\cite{mirai}, a worm-like malware that infects IoT devices and spreads by sending TCP SYN probes to pseudorandom IPv4 addresses, uses the infected devices for further scanning and DDoS attacks. Its different variants scan for vulnerabilities beyond Telnet and SSH ports~\cite{examine_mirai_sigsac2020}. 
Identifying and preventing these scans helps us limit botnet growth and nip such attacks in the bud.
CDNs also observe frequent scans, with 87\% of traffic reaching inactive ports being suspicious SYN packets indicative of scans~\cite{scanning_scanners_imc19}. A vast majority are low-volume scans originated from 13M sources, each sending fewer than 5 packets. In the context of Telnet scans of IoT devices, ~\cite{iotpot_woot15} observed that 69\% of scanners perform more than one scan, with frequent re-scanning. 
Similarly, scanning activity of Darknet IP addresses has shown continous increases from 2006-2024~\cite{internet_background_radiation_revisited_imc10,have_you_syn_me_imc24}. 
All of this give us compelling evidence that scans are a critical vector to enable many attacks, and that in aggregate they can impose a substantial load on end hosts if they are not blocked by the network earlier. 

Relegating detection of these attacks to end-hosts is fraught with risk since victim machines may be misconfigured, allowing the attacks to impact the services provided by those hosts. 
Production networks currently rely on ``Scrubbers,'' which use stateful monitoring and deep packet inspection to separate good from bad traffic~\cite{moura_into_2020}. However, Scrubbers are expensive, add latency to all traffic that must traverse them, and may themselves suffer from overload.

Programmable data planes composed of programmable switches (referred to simply as switches hereafter), smart network interface cards (\snics), and hosts can enable new ways to monitor modern communication networks, but each technology presents different trade-offs in terms of speed, memory capacity, and programmability.
In this work, we seek to combine the strengths of each data plane technology to find the proverbial ``needle-in-the-haystack'' of low volume and slow attacks, while operating at Terabit speeds. 

Our design, \name, introduces a layered defense architecture that utilizes switch-based dynamic filters, SmartNIC-based data structures and machine learning models, and cross-flow analysis for attack prevention. We `flip' the script for attack detection by focusing on how benign traffic can be processed more quickly, leaving more resources for identifying malicious flows. The latter undergo stateful analysis and deep cross-flow analysis on SmartNICs. This approach leverages novel data structures for stateful forwarding of benign traffic with minimal errors. We employ machine learning to quickly classfy normal traffic using only packet header information that is available even for encrypted traffic.

\name is designed to be deployed at the edge of data centers to act as a perimeter defense against incoming attacks. We expect \name to have visibility into traffic going in and out of the data center, and leverage a combination of data plane devices to achieve high performance.  Our contributions include:
\begin{itemize}[leftmargin=10pt,nosep]
    \item We propose a novel architecture that leverages switches and \snics to detect and mitigate low-volume and slow attacks in real-time.
    \item We introduce an approximate data structure for efficiently tracking large sets of flows that minimizes false positives, makes efficient use of memory resources, and aligns with the programming model of any P4-pipeline based switch.
    \item We develop SmartNIC-based data structures and ML models that can quickly classify flows as benign or malicious based on early packet behavior. These decisions are aggregated across flows using well-understood IDS rules to accurately identify malicious sources.
    \item We design data-plane based communication protocols and batching optimizations to efficiently coordinate rules across the switch, NIC, and host.
\end{itemize}

We evaluate our design with both trace-driven simulations and a testbed implementation using a Tofino P4 Switch and a NVidia Bluefield 3 DPU \snic. 
While our implementation of \name uses a Tofino switch, our design is fundamentally centered around multi-stage match-action rule-based switches, and should be applicable to other such switch designs.
In our experiment with a single SmartNIC core and  100 Gbps port, we achieved an end-to-end throughput of 96 Gbps, while only 0.14 Gbps was forwarded to the SmartNIC. When extrapolating up to 1.5 Tbps of traffic entering the switch, the amount of traffic that must be processed by the SmartNIC increases only slightly, to 2.24 Gbps.
Our design has high accuracy, correctly identifying 99\% of scan attacker IPs in the CAIDA trace, with orders of magnitude fewer false positives than state of the art NetBeacon~\cite{zhou_efficient_2023}.
\section{Background and Related Work}
\label{sec:background}

\subsection{Low Volume and Slow Attacks}
Slow and stealthy attacks can overwhelm or exploit individual nodes through targeted, low-intensity network activity. 
These attacks may otherwise be missed in the presence of a very large volume of traffic. We focus on the following attack types:
\emph{Port and Address Scan Attacks} are {\em low volume} methods for discovering exploitable open ports of network servers~\cite{PortScan}.
\emph{Slowloris}, a typical slow attack~\cite{archiveSlowlorisHTTP} holds connections open by continuously sending partial HTTP requests at a purposefully {\em slow} rate to keep sockets from closing.
\emph{SSH/FTP brute-forcing} is an attack where one or more nodes use different username/password combinations to try to log in to a site~\cite{ssh_bruteforce}. 
\emph{Distributed SYN Attacks} use multiple sources to send SYNs to a victim either for scanning or to exhaust a target’s kernel socket resources. 

Detecting these attacks is challenging since each flow has only a few packets and it can require stateful processing to account for timing or repeated attempts that indicate an attack rather than benign behavior. 

\subsection{Low Volume Attacks in the Wild}

To understand the prevalence of low-volume attacks in the Internet, we examined TCP flows in two publicly available traces:
1) MAWI Working Group Traffic archive~\cite{cho00:wide-data-repository,mawi-traffic-archive} collected in 2024 from the WIDE backbone in Japan, and 2) CAIDA~\cite{CAIDA_2019_trace}, collected in 2019 from an Equinix datacenter in New York, NY. 
Since both the MAWI and CAIDA traces were collected from a point in the core of the network (e.g., at an IXP), they contain some flows following asymmetric routes~\cite{routing_symmetry_iwcmc10}. 
\footnote{The degree of routing symmetry 
decreases near the core of the Internet, due to hot-potato routing and other peering artifacts. In fact "only 4-5\% of tuples generate traffic routed symmetrically"~\cite{routing_symmetry_iwcmc10}. 
Thus, only one direction of some flows may be observed, and we exclude these flows from our analysis. }
We focus on flows where the trace contains the records for a flow's packets seen in both directions, and we see the start of the flow (SYN). 

We then used the Zeek~\cite{zeek} Intrusion Detection System to classify flows as either Benign or related to one of the known attack types to provide a ground truth. Zeek's rules are somewhat conservative, so a single incomplete flow does not signify an attack -- only if a sufficient number of incomplete flows are generated from a source (or towards a victim) will it be flagged. We follow the same principle, meaning that we classify some flows as Incomplete (not necessarily part of an attack) and others as Scans (i.e., incomplete flows from a source IP that makes multiple failed attempts). We use backscatter analysis from~\cite{Flood_Moore} to detect Distributed SYN attacks. We have observed some sources, sending data packets after disregarding a server's RST response during connection establishment (SYN, SYN-ACK, RST from victim, ACK (with data from source). For these, we label them as Non-benign.

\subhead{MAWI Dec.:} Our analysis of the 15 minute MAWI trace captured at 2 pm on December 16th, 2024 in Table~\ref{tab:mawi_1216} reveals that Zeek identifies over 9.7M out of 13.7M flows as scans. Of the 62M packets in the trace, approximately 10.5M packets were from scan flows (16.9\% of the total packets), yet they originated from only 2.8\% of the source IPs. The MAWI Nov. trace, shown in Table~\ref{appendix:mawi_1115} in the Appendix, has similar behavior.

\subhead{CAIDA 2019:} We combine five 1-minute CAIDA traces from both directions of the January 17th, 2019 trace.
We find (as shown in Table~\ref{tab:caida}) the percentage of scan attack flows to still be quite significant, at 69.4\% of the total flows, while the percent of packets is even more than MAWI at 30\%. 

In both traces, we observe that scanner flows typically contain only 1–-2 packets, whereas benign flows exhibit a much larger average flow length (46–-98). We leverage this key distinction in our design: if we can quickly identify and separate out benign flows, then examining the packets of very short flows more thoroughly for attacks becomes feasible.

To increase the complexity of attacks being considered in our evaluation, we also artificially add flows for Slowloris, SSH and FTP Brute-forcing since we were not able to observe these attacks in these MAWI and CAIDA traces. These flows are created using attack generation tools and then interleaved with the existing traces for accurate timing.

\begin{table}[t]
\centering
\caption{MAWI Trace (Dec. 16, 2024) [14:00-14:15]}
\begin{small}
\begin{tabular}{ l ccc  } 
  \toprule
 Field & IPs & Flows & Packets \\
  \midrule
  Total & 620,531 & 13,700,157 & 62,105,669 \\
  Scan \% & 2.8\% & 71.3\% & 16.9\% \\
  Benign & 42,656 & 538,175 & 46,858,539 \\
  Incomplete & 430,965 & 3,285,540 & 4,004,990 \\
  Non-Benign & 7,574 & 102,232 & 1,069,150 \\
  Scan & 513,498 & 9,774,209 & 10,514,794 \\
  Distributed Syn & 1,419 & 13,230 & 45,788 \\
  \bottomrule
\end{tabular}
\end{small}
\label{tab:mawi_1216}
\end{table}

\begin{table}[t]
\centering
\caption{CAIDA Trace (Jan. 17, 2019) [13:04-13:08]}
\begin{small}
\begin{tabular}{ l ccc  } 
  \toprule
 Field & IPs & Flows & Packets \\
  \midrule
  Total & 5,526,203 & 11,079,816 & 29,285,396 \\
  Scan \% & 0.2\% & 69.4\% & 30.0\% \\
  Benign & 118,961 & 303,589 & 13,853,822 \\
  Incomplete & 859,398 & 3,085,093 & 6,913,361 \\
  Non-Benign & 6 & 3 & 38 \\
  Scan & 4,927,718 & 7,691,131 & 8,796,604 \\
  Distributed Syn & 2,688 & 7,695 & 22,120 \\
  \bottomrule
\end{tabular}
\end{small}
\label{tab:caida}
\end{table}

\subsection{Low and Slow Attack Mitigation}

Unlike heavy attacks based on long running flows, the analysis in the preceding section shows that scans and slow attacks arise from a large number of short connections, many recurring from the same source IPs. Blacklisting IPs is a common solution to such attacks~\cite{reading_tea_leaves_usenixsecurity2019, BLAG_NDSS2020}, and there are public lists for known malicious IP ranges~\cite{iblocklist, snort, nothink, BruteForceBlocker, opendbl, dshield, badip}. However, dropping all packets from an IP can have negative effects on legitimate traffic, particularly in the presence of Network Address Translation (NAT)~\cite{nat_blocklist_imc2020}. Instead, it is preferable to have traffic from suspicious flows carefully analyzed by a scrubber running intrusion prevention software.

Unfortunately, state of the art IPS platforms struggle to achieve the throughput needed for high speed networks. Pigasus supports up to 100Gbps on one host using FPGA accelerators~\cite{pigasus}, but it reports that a CPU-only approach could require more than 500 cores. This motivates the design of \name, to efficiently separate out which IPs are sending traffic that may be suspicious so they can be analyzed more deeply, while still providing security for all flows. 

\subsection{Data Plane-based Monitoring}

Existing methods detect volumetric attacks at Terabit speeds using approximations like sampling and sketches~\cite{gupta_sonata_2018,BeauCoup,liu_nitrosketch_2019,tang2019mv,Elastic18}, but struggle with accuracy on non-heavy hitter flows and cannot maintain complex state. Solutions like Elastic Sketch~\cite{Elastic18} and SketchVisor~\cite{SketchVisor17} lose fine-grained accuracy as flow counts grow, while Nitrosketch~\cite{liu_nitrosketch_2019} and UnivMon~\cite{liu_one_2016} improve accuracy but still miss many mice flows due to sampling.

Recently, ML-based techniques have been proposed to identify attack traffic~\cite{jafri24:leo,zheng_iisy_2024,BoS}. NetBeacon~\cite{zhou_efficient_2023} deploys ML packet and flow based models to a switch to classify traffic at line rate. But our evaluation illustrates that even at modest traffic rates, a switch-based flow state cache can become overloaded, causing the system to revert to using the less accurate packet-based classifier. Since this is stateless, it may be unable to detect more complex attacks.
IIsy~\cite{zheng_iisy_2024} and Leo~\cite{jafri24:leo} deploy the decision trees or random forests on a switch,  but they are limited in the list of features they support due to memory limits on the switch. Brain-on-Switch~\cite{BoS} proposes an RNN model on the switch and uses a transformer on the host. However, having the transformer on the host incurs delays of the order of seconds, which is undesirable.

\subsection{Heterogeneous Data Plane Devices}
\label{sec:heterogeneous-data-plane}

This table characterizes resources and performance of heterogeneous data plane devices available in our testbed. 

\begin{small}
  \begin{tabular}{@{}llll@{}}
    \toprule
    & Throughput  & Memory  & Programmability \\ \midrule
    Switch (per pipe) & 1.6 Tbps    & 6.9 MB & P4 \\
    NIC (per port)    & 200 Gbps     & 32 GB    & C \\
    Host (per core)   & 10--24 Gbps     & 25 GB   & All \\ \bottomrule
  \end{tabular}
\end{small}

\subhead{Switches:} A Tofino v1 switch typically has two or four pipes, being able to handle an aggregate forwarding rate of up to 6.4 Tbps, with guaranteed line rate performance due to the strict pipeline based programming model. However, even the four pipeline switch contains only 24 MB of memory for stateful processing, and not all of it can be used by dataplane programs. According to \cite{tofinoforum}, the available memory for registers in a pipe can be up to 6.9MB. 

Although switches have been used for traffic monitoring \cite{BeauCoup,gupta_sonata_2018,barradas21:flowlens,xing20:netwarden,zhang2020gallium}, their limited SRAM memory for the program makes it difficult to retain state across packets or to hold the exact-match tables~\cite{barradas21:flowlens}. Previous work \cite{hill_tracking_2018,zhou_cerberus_2024}, has also shown that due to resource limitations, standalone switch solutions will suffer under high traffic, and defensive mechanisms can still be easily exploited. It requires the monitor to focus on subsets of the traffic and use dynamic iterative query refinement processed at the switch~\cite{gupta_sonata_2018} to hone-in on the correct traffic subset that contains attack traffic.

\subhead{\snics:} 
The NVIDIA BlueField 3 \snic that we use has two 200 Gbps ports and a total of 32 GB of DRAM~\cite{nvidia-bluefield-3-spec}. While this card provides greater programmability than the switch, it's processing capacity is much lower.

\snics have been used for packet traffic monitoring~\cite{panda_smartwatch_2021,sonchack_turboflow_2018,pigasus}, but they have limited bandwidth capacity, and their processor architectures do not guarantee line rates like switches do. Although they may outperform host-based solutions~\cite{sonchack_turboflow_2018, pigasus}, effective traffic monitoring at Terabit speeds would require a large number of such \snics. Despite these limitations, \snics can still nicely complement switches with their larger memory. For example, DeepMatch \cite{hypolite_deepmatch_2020} and Synergy~\cite{panda_synergy_2022} enable architecture-aware style programming of \snics to support large traffic volumes, and iPipe \cite{iPipe19} helps reduce host CPU load. 

\subhead{Hosts:} Network Function Virtualization software has developed ways for hosts to very efficiently process packets at high rates and low latency, often by bypassing the kernel and using shared memory to eliminate copying~\cite{martins_clickos_2014,palkar_e2:_2015,rizzo_netmap:_2012,zhang_opennetvm:_2016}.
However, each core on the host running high performance DPDK software still can only handle about 10 Gbps when processing small packets (about 24 Gbps with large packets). Thus, processing terabit scale traffic would require a large number of hosts using up most of their cores.
\begin{figure*}[bt]
  \centering
  \includegraphics[width=0.95\linewidth]{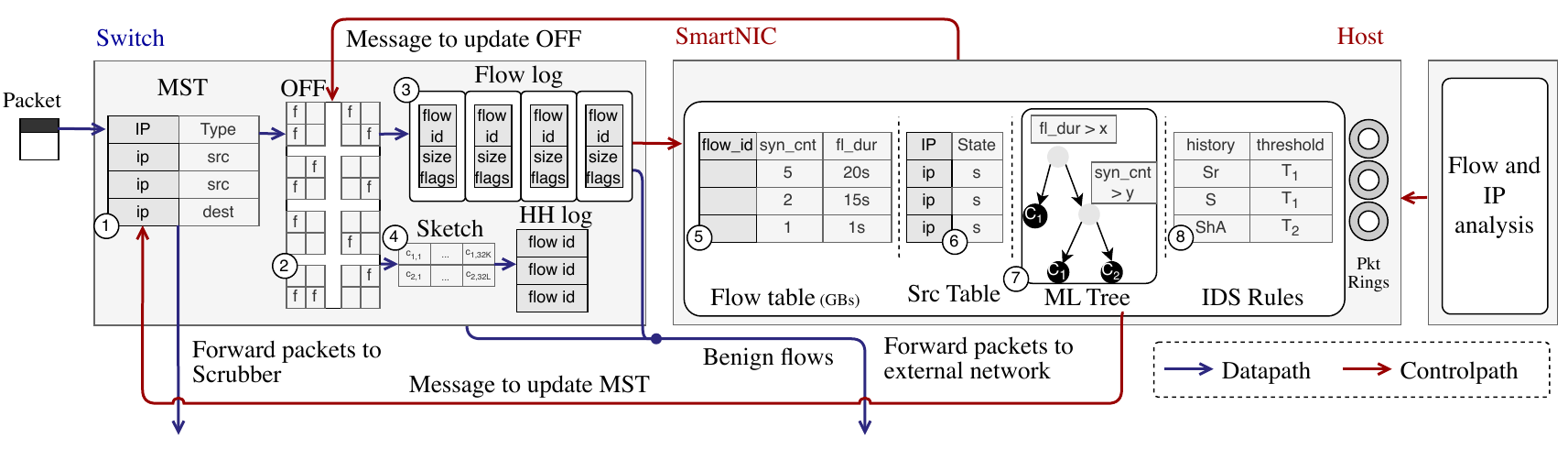}
  \caption{\name comprises:
  1) a Malicious Source Table (MST) to redirect known malicious traffic sources to a scrubber for deeper analysis;
  2) an Overwriting Flow Filter (OFF) to forward benign traffic without further analysis;
  3) a Flow Log to optimize state transfer to the \snic;
  4) a Sketch and log for tracking heavy hitters;
  5) a \snic table to track flow state ; 
  6) source table to aggregate across flows;
  7) ML classification algorithms to identify benign flows; and
  8) intrusion detection rules to identify malicious sources.
  }
  \label{fig:full_arch}
  \vspace{-3mm}
\end{figure*}

\section{Unified Switch/SmartNIC Design}
\label{sec:design}

This section presents the design of \name, a unified in-network filtering framework for identifying sophisticated and low-volume attacks hidden within terabit-scale traffic.

\subsection{Design Rationale and Overview}

The primary design principle behind \name is the division of labor between switches and \snics to leverage their complementary strengths and compensate for their weaknesses. We focus on the following characteristics: switches excel at forwarding packets at line rate by using a match-action pipeline to customize flow processing; however, their programmability and memory capacity are limited. In contrast, \snics offer greater programmability and larger memory capacity, but their bandwidth is significantly lower.

To exploit and compensate for these characteristics, \name assigns complementary roles to the switch and the \snic. The primary objective of the programmable switch is to reduce traffic to the \snic. To achieve this, the switch performs stateless and space-efficient filtering, which redirects flows that have already been identified as either benign or malicious at terabit-scale line rate. 

In contrast, the primary objective of the \snic is to accelerate the control loop of \name, i.e., capture traffic, classify it as benign or identify it as malicious, and inject updated rules into the switch. Specifically, because the \snic resides between the switch and the host, it can shorten the control loop by identifying part of the malicious and benign flows locally.
To achieve this, the \snic implements a lightweight ML-based classifier combined with a rule-based flow classification mechanism. Benign flows are quickly classified by the ML-based classifier, while other flows require rule-based identification with per-flow and cross-flow features to determine the nature of the attack. 

This combination enables \name to rapidly forward benign traffic while identifying a malicious source that contributes a substantial fraction of malicious flows (e.g., scans) using relatively simple rule patterns, as investigated in \S~\ref{sec:background}.
By offloading these decisions to the \snic, \name can operate effectively at terabit scale while preserving accurate detection of sophisticated and low-volume attacks.

Figure~\ref{fig:full_arch} outlines the architecture of \name. The design begins with a malicious source table that redirects traffic identified as suspicious to a scrubber for deeper analysis (described in \S~\ref{sec:switch-mst}), while other flows are forwarded to a benign packet filter that directly forwards packets from known normal flows (\S~\ref{sec:switch-off}) towards the destination. Only unknown traffic is then passed to the \snic for rapid classification or identification. We deploy a lightweight ML classifier to detect benign flows (\S~\ref{sec:snic-benign}), complemented by attack-specific rules adapted from existing IDSs, such as Zeek, to identify malicious sources (\S~\ref{sec:snic-zeek}). Finally, the \snic feeds results to the switch to continually refine filtering accuracy (\S~\ref{sec:switch-snic}).

\subsection{Switch: Malicious Source Filtering}
\label{sec:switch-mst}

The \name system observes the behavior of sources of attacks, allowing it to reduce the load for \snic based analysis by efficiently applying a blacklist of malicious source IPs. The blacklist is determined by the \snic, by aggregating information across multiple attack flows over a slightly longer timescale.
The switch maintains this Malicious Source Table (MST) indicating  IP addresses that have been associated with malicious flows.
This list of sources can be effectively tracked using the switch's TCAM, which allows either exact match or range-based match rules to be defined. The MST tracks the sources of attacks such as scans and slowloris. It also tracks the destination for Distributed SYN attacks where many sources may target one victim. While our primary focus is slow attacks, we also demonstrate how \name could also track heavy flows (e.g., high volume attacks).

Since the MST is stored in the switch TCAM, it has a limited amount of memory available, and it can only be updated by the switch's relatively slow control plane. As we evaluate later, we find that this is sufficient for our purposes---while the number of malicious flows can be very large, many typically come from a single source. This reduces both the memory size needed to track the flows, and the frequency with which they must be updated.

We choose to forward packets that see a match (`hit') on the MST to a scrubber for further analysis, rather than dropping them. There are two considerations that prompt this design. Even if there is a relatively small false positive of packets of benign flows seeing a hit on the MST, dropping these packets is undesirable. Moreover, in the presence of Network Address Translation (NAT), the IP address visible for \name to populate the MST may be the NAT source address. 
As we observe in the December MAWI trace, the top 2 heavy hitter (having most packets) source IPs are both labeled as Address Scans by Zeek. The largest source IP produces 35,697 benign flows generating 3,566,038 benign packets but also has 58 flows labeled as address scans contributing 309 packets. The second largest source IP shows a similar pattern, with 29,170 benign flows generating 3,160,570 packets while only 78 flows (labeled as address scans) contributed 502 packets. 
The scrubber can classify the flows more carefully (i.e., using more cross-flow state and deeper examination of the flow and packet features (e.g., payload length) by buffering and examining more packets of a flow). The scrubber may then adjust this policy via the switch control plane so packets of such benign flows may then be forwarded towards the destination, avoiding the potential of adding more delay. 

\subsection{Switch: Benign Flow Filtering}
\label{sec:switch-off}

\begin{figure}[t]
    \centering
    \begin{subfigure}[b]{0.49\linewidth}
         \includegraphics[width=\linewidth]{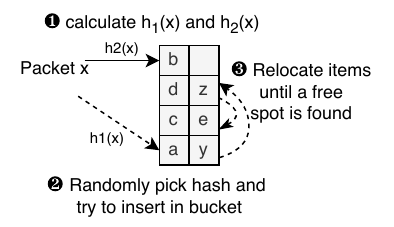}
         \caption{Cuckoo Filtering}
        \label{fig:cuckoo}
    \end{subfigure}
    \hfil
    \begin{subfigure}[b]{0.49\linewidth}
        \centering
        \includegraphics[width=\linewidth]{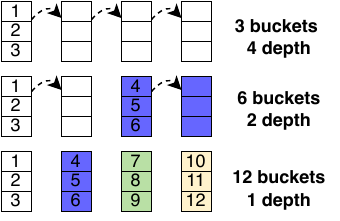}
        \caption{OFF Table Layout}
        \label{fig:off-layout}
    \end{subfigure}
    \caption{(a) Cuckoo Filtering requires access to multiple entries within two buckets to do lookups and insertions, which would require many recirculations if implemented on a switch. (b) Our OFF structure places one register array in each pipeline stage; with four stages, these can be structured to maximize depth or number of buckets.}
    \label{fig:filters}
\end{figure}

Tracking the millions of benign flows present in Terabit rate networks necessitates an efficient whitelisting method to track flows that don't need further analysis. Bloom Filters~\cite{broder_network_2002} and Cuckoo Filters~\cite{fan2014cuckoo} offer memory-efficient set membership checks, but are challenging to implement on P4 switches. Bloom Filters become saturated over time, reducing accuracy. Cuckoo Filters support flow removal, but identifying expired flows in the network monitor is complex, if not infeasible. 
Figure~\ref{fig:cuckoo} illustrates how an insert into a Cuckoo Filter proceeds by using two hash functions to generate possible buckets (rows), one of which is picked at random to store the new item's fingerprint. If all entries in the bucket are full, then an existing fingerprint is relocated to a different bucket by calculating its secondary hash. This is repeated until an empty entry is found. While this allows the Cuckoo Filter to do fast lookups even with high occupancy, its access pattern conflicts with the requirements of a P4 pipeline.

Memory on a typical P4 switch is divided into register blocks, and access to them is highly constrained: 1) each register block is only accessible by a single stage in the pipeline to prevent race conditions; 2) only one entry per register block can be read per pipeline stage; 3) data that is read from a register is not available until the end of the stage it was accessed in, preventing conditional updates; 4) the total number of pipeline stages is limited by the hardware (e.g., 12 on our Tofino v1). 
Thus the structure of the Cuckoo Filter violates many of these properties, e.g., a Cuckoo Filter lookup requires checking all cells in both buckets selected by the hash functions, but each of these accesses will need to be performed in different stages; the relocation process, critical to the Cuckoo Filter's high utilization, can require up to 500 iterations in a previous implementation~\cite{fan2014cuckoo}, but this will incur unreasonable overhead from excessive recirculations in a P4 pipeline.
Moreover, in our context, it is important to focus on reducing false positives (incorrectly identifying malicious flows as benign so they are forwarded without further examination) since these can be more harmful than false negatives, yet both the Bloom and Cuckoo Filters focus on the opposite.
Consequently, a specialized design is needed that minimizes memory usage while carefully managing the trade-off between false positives and the operational overhead of false negatives.

\head{Overwriting Flow Filter (\off):}
We present a new filter data structure that builds on the design philosophy of Cuckoo filters.
Instead of assuming the presence of empty cells, we design our Overwriting Flow Filter (\off) like a cache that is always completely full, but where entries will be evicted (overwritten) over time.
We carefully design our flow filter data structure to circumvent a programmable switch's memory and stage constraints, and realize constant-time entry insertion and lookup with no recirculations.

\subhead{Insert and Lookup Algorithms:} We structure the flow filter as an array of buckets, $f_i[\:]$,  where each entry in a bucket contains a small fingerprint indicating the signature of that flow. The filter is configured by these parameters:
$m$, the number of buckets in the array;
$b$, the number of entries in each bucket; and
$f$, the length of the fingerprint in bits.
When a flow has been deemed safe and is to be inserted into the filter, the switch performs the following procedure:
\begin{enumerate}[leftmargin=10pt,nosep]
  \item Calculate hash and fingerprint: The packet's 5-tuple is hashed to determine the index for the buckets that will be used as $i = h(Pkt) \: \% \: m$. A $f$-bit fingerprint, $x$, is also calculated to represent the packet.
  \item Store fingerprint: The fingerprint is saved in the first  entry of bucket $i$, e.g., $f_i[0] = x$.
  \item Demote and evict: Any existing entries in the bucket selected for insertion 
  are demoted to the next level so the cell can be overwritten. This is repeated, with the cell in the last bucket entry being deleted.
\end{enumerate}
In order to perform a lookup to test if an element has been stored in the flow filter we follow a similar procedure and check if any of the entries in the bucket indicated by the hash function match the fingerprint.

Unlike Cuckoo Filters, we use a single hash function; using multiple hash functions is effective when a table is partly empty (since there is a chance one of the selected buckets will have an empty cell), but that will not occur in our design at steady state where all entries are occupied. Rather than rehash and relocate entries on a collision, we perform demotion within the bucket.
This makes our structure mimic a FIFO eviction policy.\footnote{We also attempted a LRU variant, but found that the potential improvement in hit rate was outweighed by the cost of recirculations. We settle on this simpler structure as it fits well within constraints of the switch's pipeline.}

\begin{figure}[t]
  \centering
  \includegraphics[width=0.99\linewidth]{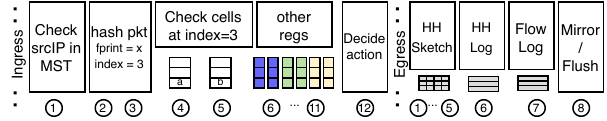}
  \caption{\name Pipeline Stages}
  \label{fig:pipeline}
\end{figure}

\subhead{Flow Filter Accuracy Metrics:} The \off data structure has the possibility of both false positives and negatives. A {\em false positive} occurs when the fingerprint for the packet that is being queried collides with one stored in {\em any} of the $b$ entries in the bucket checked in the lookup. To calculate the probability of such a collision, we assume the worst case where all cells have had a fingerprint stored in them. For a fingerprint size of $f$ bits, this gives false positive probability: $\mathit{FPR} = b/2^f$.

This equation is derived by considering the probability of any of the $b$ fingerprints in a bucket colliding (each with probability $1/{2^f}$), assuming uniform random fingerprint distribution.
Thus increasing the number of entries in a bucket raises the chance of false positives (albeit with other benefits discussed later).
In our implementation we set $f=16$, and $b=2$, giving a false positive rate of $0.003\%$, or about 1 in 33 thousand flows. We believe this is low enough to thwart most attackers since each attempt at a collision requires a different 5-tuple and we salt the fingerprints so that malicious users cannot predict which values will cause a collision.

Calculating the {\em false negative} rate is more complex since it depends on the likelihood of new safe flows arriving with the same hash values, causing prior flows to be evicted. \off evicts the oldest flow in each bucket during an insertion, if that flow is still active, then the next time a packet arrives for it there will be a false negative. This will result in the packet being forwarded to the \snic/host for analysis, which will re-insert the flow into the filter as safe. 

\subhead{Structure and Switch Limitations:}
Given a fixed memory size available on the switch, there is a trade-off between the number of buckets, $m$, and the number of cells within each bucket, $b$. This is illustrated in Fig~\ref{fig:off-layout}, where a filter with 12 total entries has different layouts. The top example uses ($m=3$, $b=4$), leading to multiple chances for a fingerprint to be demoted before eviction (helping false negatives), but a higher chance for an initial hash collision (hurting false positives). The bottom example has the other extreme of ($m=12$, $b=1$); based on our false positive equations and simulation results described later, we select $b=2$, and maximize $m$ based on the available memory and stages.

To meet the P4 pipeline memory access requirements, we structure the OFF as a series of registers that are subdivided across the switch's pipeline stages---entries from each bucket must be stored in different registers in order to allow demotion without recirculation.
Our implementation on a Tofino v1 switch uses eight memory registers, each with $2^{18}$ 16-bit cells. Registers from each pair of stages are treated as arrays of buckets with depth two.
In total, this structure allows us to store a maximum of $2^{18} \times 8 \approx 2$M fingerprints using 4 MB of switch memory, spread over eight pipeline stages.
If the switch had additional memory and stages, then the register size could be increased.

As shown in Figure~\ref{fig:pipeline}, the first few stages of the pipeline are used to check the MST and then calculate the 16 bit fingerprint and a 32 bit hash value for the packet.
The hash is used to select the bucket and determine the stage where the bucket's first entry is stored. In the example, $index=3$ maps to the third bucket, which is spread across stages 4 and 5. In this case a lookup for fingerprint $x$ would fail since it doesn't match $a$ or $b$ stored in the bucket's entries; for an insertion, $x$ would replace $a$ in stage 4, with the overwritten value demoted to replace $b$ in stage 5, which would in turn be evicted. Stages 6-11, holding other register blocks not used in this lookup, would not be accessed. The final stage 12 of the ingress pipeline is used to determine if the packet can be immediately sent out (a hit in the \off) or if it must be logged for later transmission to the \snic (Section~\ref{sec:switch-snic}).

\subsection{Switch: Heavy Hitter Detection}

While our primary focus is on detecting low volume and slow attacks, we recognize that such a system will also benefit from detecting high volume attacks or anomalies such as heavy hitters. \name can support this by integrating sketch techniques~\cite{namkung_sketchlib_2022,SketchVisor17,huang_omnimon_2020,liu_nitrosketch_2019} into its analysis pipeline. 

If the system wishes to detect sources generating a large number of \textit{flows}, then this can easily be accomplished by deploying a sketch data structure on the \snic, since it can increase a counter at the start of each flow. However, if the goal is to detect sources sending long running flows with a large number of \textit{packets}, then the sketch must instead be run on the switch since we need to increment counters for every packet that arrives.

We focus on how \name's design can improve the performance of a sketch. Since heavy hitters are long running flows, we only insert packets which hit in the \off into the sketch. This reduces the ``noise'' caused by packets from incomplete flows, reducing the false positive rate of the sketch as we demonstrate in our evaluation.

Deploying sketches alongside the OFF table may exhaust switch memory and stage limits. To limit this impact, we deploy our heavy hitter sketch in the egress pipeline (Figure~\ref{fig:pipeline} egress stages 1--5). However, this also complicates how we can propagate information from the heavy hitter table to the \snic or the switch's control plane, since such decisions need to be made prior to the start of the egress pipeline. To overcome this challenge, we design a heavy hitter IP log (egress stage 6) to track information that is periodically flushed to the \snic as described in Section~\ref{sec:switch-snic}. 

\subsection{\snic: Stateful Flow Tracking}
\label{sec:snic-coordination}
The \snic on the host tracks per-flow features (e.g., flow length and handshake status) and cross-flow features (e.g., the number of flows per source host) for unknown flows until they can be classified as benign or identified as malicious. This design is motivated by the fact that sophisticated attacks, such as slow scans or slowloris, 
often manifest across multiple flows. To capture such attacks, the \snic aggregates features across flows for more accurate detection.
The information is maintained in DRAM of \snic and the host utilizing a \emph{flowcache} data structure~\cite{panda_smartwatch_2021}, where the DRAM of \snic serves as the cache, called the \emph{primary buffer}, and the evicted entries are stored in host memory. The buffer uses LRU for evictions, which is effective in handling a large number of relatively short flows.

Upon receiving a flow log from the switch (the log is a batch of packet information), the host subsystem stores each flow as an entry in a table on the \snic. The \snic memory can be tens of GBytes, potentially supporting more than 100M flow entries, far more capacity than a switch.

\subsection{\snic: Classifying Benign Flows}
\label{sec:snic-benign}

The \snic uses ML to rapidly classify benign flows that do not require deeper analysis. It classifies flows into two categories: \emph{benign} and \emph{unknown}. The first pair of packets of a flow generally do not provide sufficient information for accurate classification. Once the third packet of a flow arrives (the first payload-carrying data packet of a TCP flow), the classification process can be triggered. If a flow is classified as benign, the switch's OFF (benign flow) filter is updated with an entry corresponding to the flow's five-tuple, and the corresponding flow entry is evicted from the primary buffer on the \snic.
Note that flows consisting of fewer than three packets are too short to be accurately classified using only per-flow features. Therefore, such flows are processed using rule-based detection with cross-flow features.

The goal of the ML classifier is to accurately classify benign flows while maintaining high processing speed. To this end, we adopt Leo~\cite{jafri24:leo}, a method that enables real-time packet classification through efficient decision-tree execution. Further details on model training are in the Appendix.

We tested the performance of our ML Inference, using the MAWI dataset from Dec. 2024 for training, and the Nov. 2024 dataset for testing. We achieve 93\% precision and 98\% recall when trying to classify traffic as being benign, further we find that the missclassification rate from malicious to benign is low as we demonstrate in our evaluation. 

\subsection{\snic: Identifying Malicious Flows}
\label{sec:snic-zeek}

\name leverages the vast amount of experience of IDS/IPS systems, like Bro~\cite{bro}, Zeek~\cite{zeek}, and Corelight~\cite{corelight_slowloris}, to identify malicious flows and the IP sources responsible for them.
We adapt these rules to run on the \snic, using the state from our flow table. See Appendix~\ref{sec:online-detection} for pseudo-code.

The IDS rules that we deploy are first used to identify which flows match the characteristics of attack traffic. Second, the rules apply thresholding across multiple flows to determine which sources we can confidently say are sending malicious traffic that must be scrubbed. For example, we track when the number of failed connection attempts (i.e., just a SYN or SYN + RST) surpasses a threshold to determine if a source is sending scanning traffic (Zeek suggests 25 failed connections in 15 minutes)~\cite{PortScan}.
\footnote{
Different studies use different thresholds: ~\cite{scanning_scanners_imc19} used a threshold of a source IP scanning at least 100 distinct destination IP addresses over 5400 seconds, while ~\cite{have_you_syn_me_imc24} uses multiple thresholds to understand ranges of scanners scanning from 100 ports up to more than 10,000 ports.} 
Using Zeek as a guide, we also develop attack detection scripts for Distributed SYN attack, SSH/FTP Bruteforce, and Slowloris attacks.

\subsection{Switch \snic Integration}
\label{sec:switch-snic}

Finally, our design unifies the switch and the \snic into a coordinated processing pipeline that achives line-rate processing and precise attack detection.

\head{Switch $\to$ \snic (Flow and HH Log Batching):}
\name uses the \snic to analyze packets and classify flows. With a straightforward design, every packet unknown to the switch must be sent to the \snic. To mitigate the communication cost, we develop two optimizations: \emph{packet truncation} and \emph{packet batching}.

Rather than directly forward each packet in full, the switch truncates packets to just include key flow information from the packet header needed for analysis. This can dramatically reduce the volume of data arriving at the \snic and host. This approach is inspired by seminal studies~\cite{goswami20:parking-packet-payload,scazzariello23:high-speed-stateful,yoshinaka24:p4qrs} that propose splitting headers and payloads at the switch to reduce communication and processing costs.

To further cut the number of packets sent to the \snic, the switch maintains a \emph{flow log} that buffers flow information from multiple packets and transmits it as a single batched packet only when the flow log becomes full. Even a small buffer capable of storing just a few packets yields substantial savings by eliminating redundant protocol headers and avoiding unnecessary data caused by frame padding.
With packet truncation, each flow log entry is reduced to 14 bytes: 13 bytes for the flow's 5-tuple and one byte for auxiliary information required by the \snic classifier, such as packet length and selected TCP flags. Up to three flow log entries can be packed into a single 64-byte frame.

We use a similar technique to propagate the results of the Heavy Hitter sketch to the \snic. When a packet hits in the sketch and the counter is above a threshold, we record the packet tuple in a second log in the egress pipeline. This log can then be periodically flushed to the \snic along with the log of packets that need to be analyzed.

These operations are implemented entirely in the egress pipeline, as shown in Figure~\ref{fig:pipeline}, because they are independent of routing decisions.

\head{\snic $\to$ Switch (Populating Benign Flows in OFF):}
Efficient real-time communication between the host and the switch is essential for low-latency OFF table updates. Because rule insertion via the control plane is too slow~\cite{stubbe23:keeping-up-to-date}, we adopt a data-plane-based scheme that improves both latency and throughput by several orders of magnitude. In this design, the \snic sends table update messages directly into a designated switch data-plane port, where packets arriving from the \snic port are identified to trigger OFF rule insertions, thereby eliminating control plane delays.

\head{\snic $\to$ Switch (Populating Malicious Source IPs in MST):}
Once the \snic subsystem has confidently identified a group of flows as malicious, the corresponding attacker source IP, or the destination IP of a flood victim is populated in the MST on the switch. Subsequent traffic from/to that host will be redirected by the switch to a scrubber for deeper inspection. 
Unlike benign flow updates, MST insertions occur on a much slower timescale because they require confirmation over multiple flows from a malicious source. Therefore, the MST is updated through the switch's control plane by inserting TCAM entries, which we consider acceptable given the lower update frequency.

\section{Data-Driven Justification of Design}
\label{sec:justification}

Here we analyze how our switch-based filtering architecture can achieve high accuracy monitoring of terabit-scale traffic.

\subsection{Implementation}
\label{sec:implementation}
\begin{figure}[t]
  \centering
  \includegraphics[width=0.9\linewidth]{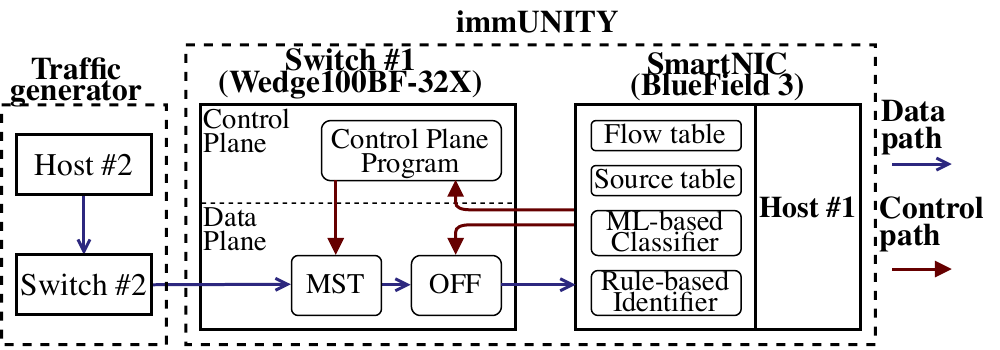}
  \caption{The \name testbed implementation}
  \label{fig:prototype}
\end{figure}

We implement \name on a Tofino~1 switch (Wedge 100BF-32X) and a Bluefield~3 \snic (B3220), as shown in Fig.~\ref{fig:prototype}.
On the switch, we allocate 4 MB of registers to the \off filter (2M entries) and 40 KB of SRAM to the MST (10K sources). We deploy a sketch in the egress pipeline based on UnivMon~\cite{liu_one_2016}. Due to switch memory limits we use 2 hash functions (rows) in the sketch, each with 32K 4-byte counters.
For fast MST updates, we build a control plane program that receives control packets for MST updates via Tofino's internal port, which connects the data plane to the control plane. This internal port is implemented over the PCIe bus and enables packets to be transferred from the data plane directly to our user-space control-plane program, bypassing L3/L4 processing in the control plane OS kernel.

We allocate 17 GB of memory for the flow tables on the \snic, which accommodates up to 110M flow entries. For cache management, we adopt Bubble LRU~\cite{zhang09:bubble-LRU}, which is an efficient approximation of true LRU and significantly reduces update overhead with similar eviction behavior.

\subsection{System Capacity for Attack Detection}
We first demonstrate that our design ensures incoming traffic does not overwhelm any processing component. Specifically, for each component $x$, the packet arrival rate satisfies $\lambda_{x} \le \mu_{x}$, and for each link between components $x$ and $y$, the traffic load remains within the available bandwidth $B_{x,y}$, even when the total traffic volume scales to Tbps levels.

\begin{figure*}[t]
  \begin{minipage}[b]{0.7\textwidth}
    \centering
    \includegraphics[width=1.0\linewidth]{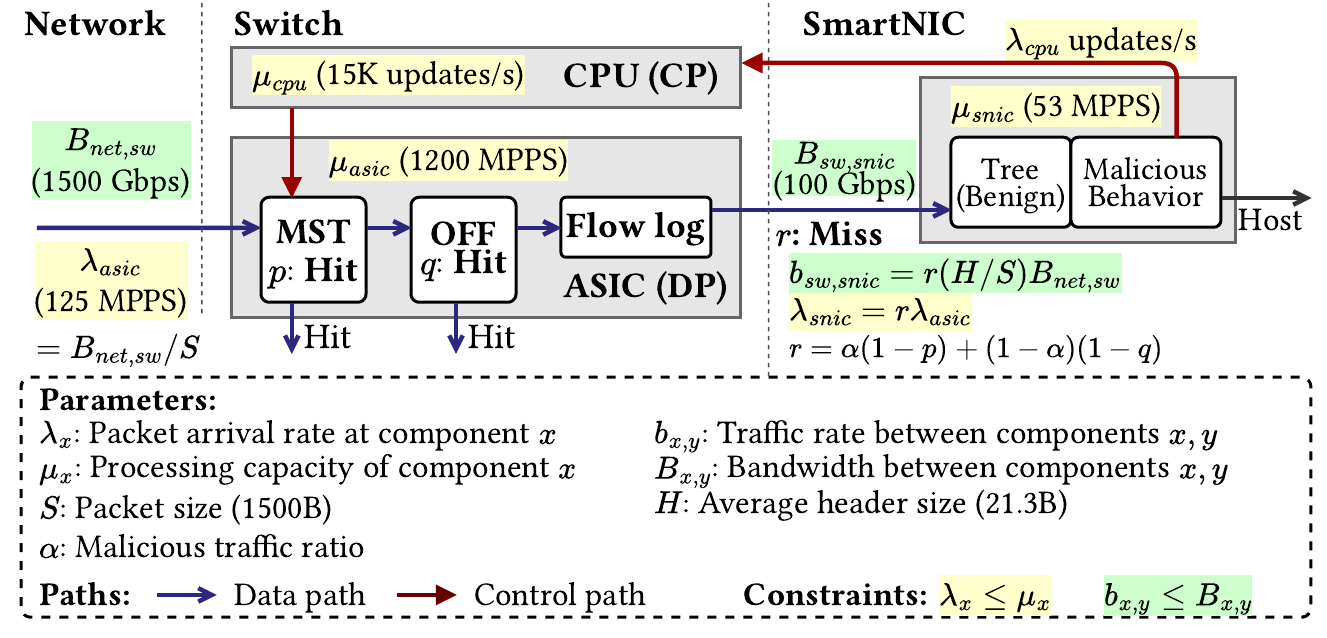}
    \vspace{-4mm} 
    \caption{The system model for analyzing \name. The equations highlighted in green represent values in bits/s, while those highlighted in yellow indicate values in packets/s.}
    \label{fig:system-model}
  \end{minipage}
  \hfil
  \begin{minipage}[b]{0.27\textwidth}
    \centering
    \includegraphics[width=1.0\linewidth]{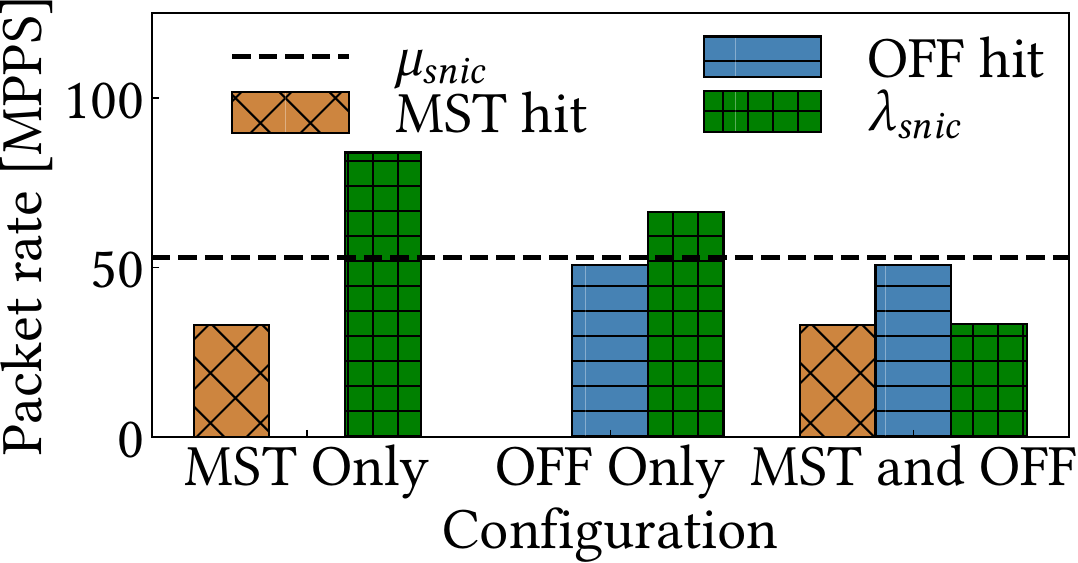}
    \vspace{-4mm} 
    \caption{Load on MST, OFF, and \snic for 3 configurations}
    \label{fig:snic-load}
    \vfill
    \includegraphics[width=1.0\linewidth]{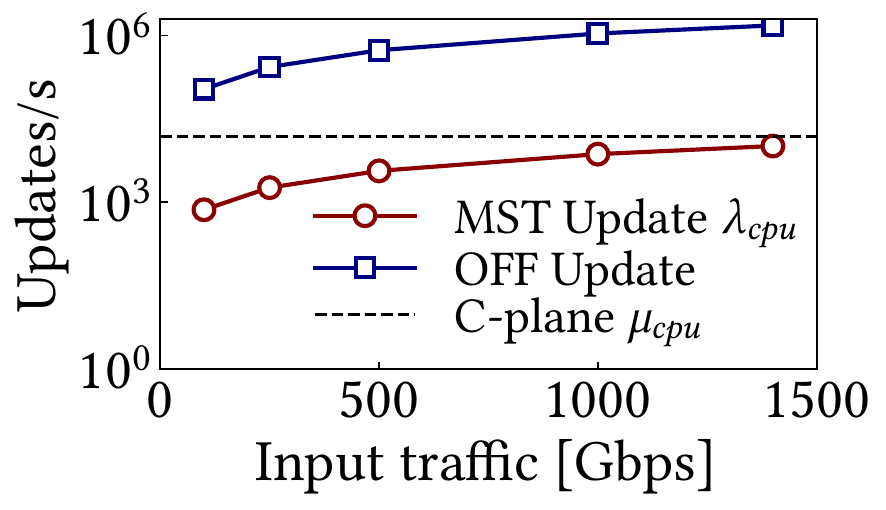}
    \vspace{-4mm} 
    \caption{MST and OFF update rate}
    \label{fig:update}
  \end{minipage}
  \vspace{-3mm}
\end{figure*}

\subsubsection{Modeling Loads and Capacities of Components:}
Figure~\ref{fig:system-model} illustrates the system model assumed in our analysis. 

Incoming traffic arrives at a rate of $B_{\mathit{net},\mathit{sw}}$ Gbps, or $\lambda_{\mathit{asic}}$ MPPS in packet units. This traffic consists of malicious traffic at rate $\alpha \lambda_{\mathit{asic}}$ and benign traffic at rate $(1 - \alpha)\lambda_{\mathit{asic}}$. Let $p$ and $q$ denote the hit rates of the MST and OFF tables, respectively. A hit in the MST filters out a fraction $\alpha p$ of the malicious packets (which are not forwarded to the destination but are instead sent to a scrubber). Likewise, a hit in the OFF filter causes a fraction $(1 - \alpha) q$ of the benign traffic to bypass further processing and go directly to the destination. Hence, the fraction of traffic forwarded to the \snic is
\begin{equation}
    r=\alpha(1-p)+(1-\alpha)(1-q),
    \label{eq:r}
\end{equation}
and the resulting packet arrival rate at the \snic is
\begin{equation}
    \lambda_{\mathit{snic}} = r \lambda_{\mathit{asic}},
    \label{eq:lsnic}
\end{equation}
where $\alpha(1-p)$ is the malicious traffic missed by the MST and $(1-\alpha)(1-q)$ is the benign traffic missed by the OFF.

We truncate and batch packets sent to the \snic. Truncation reduces the size of a packet from $S$ bytes to a header of only $H$ bytes, lowering the traffic between the switch and the \snic ($b_{\mathit{sw},\mathit{snic}}$) by a factor of $H/S$. Accordingly, we estimate the bandwidth consumption as
\begin{equation}
  b_{\mathit{sw},\mathit{snic}} = r \times H/S \times b_{\mathit{net},\mathit{sw}}.
  \label{eq:bsw-wnic}
\end{equation}
Our batching mechanism further reduces the number of packets sent to the \snic by a factor of 3. However, we conservatively ignore this optimization here because the \snic must still perform four state table updates for each incoming packet. As a result, the computational load on the \snic CPU remains similar, although link bandwidth consumption and I/O overhead are reduced.

Finally, the \snic requests the host CPU to perform an update to the MST whenever it identifies a new malicious source based on the Zeek rules. The update rate is denoted as $\lambda_{cpu}$ updates/s in the figure. The \snic also sends updates to the OFF via the dataplane, so it must be below the bandwidth of a single port.

\subsubsection{Obtaining Actual Loads and Capacities:}
We next derive the capacity of each component based on hardware specifications and empirical measurements with our implementation.

The switch uses one pipeline of the Tofino ASIC to implement \name, yielding a processing capacity of $\mu_{\mathit{asic}} = 1200$ MPPS~\cite{intel:tofino}. One of the switch's 16 ports is used to connect to the \snic via a 100 Gbps interface ($B_{\mathit{sw},\mathit{snic}} = 100$ Gbps), while the remaining 15 100-Gbps ports accommodate incoming traffic ($B_{\mathit{net},\mathit{sw}} = 1500$ Gbps). Based on the MAWI Dec.\ trace statistics~\cite{mawi:statistics-20241216}, where TCP packets have an average size of 1737 bytes, we approximate the packet size as $S = 1500$ bytes in our analysis. Given this packet size, the switch processes traffic at approximately $\lambda_{\mathit{asic}} \approx 125$ MPPS. We also experimentally measure the update rate of the TCAM-based match-action table using the control plane CPU, and we obtain an update rate of $\mu_{\mathit{cpu}} = 15,385$ updates/s.

For the \snic parameters, we measure a single-core processing rate by implementing a flow table (8 M entries), a decision tree, and Zeek scan-detection rules on one Arm core on the \snic, and obtain a processing rate of 3.3 MPPS. Assuming linear scaling across 16 Arm cores, we estimate a total \snic processing capacity of $\mu_{\mathit{snic}} = 53$ MPPS; the full multi-core implementation is currently in progress.

\subsubsection{Traffic Reduction by Flow Logging:}
We first examine how the switch's flow log reduces the amount of traffic sent to the \snic ($b_{\mathit{sw},\mathit{snic}}$). 
Each packet is truncated to $H = 21.3$ bytes since our implementation extracts and batches information of three packets into a single 64 bytes packet. This results in $H/S = 0.014$.
According to Eq.~\eqref{eq:bsw-wnic}, this implies $b_{\mathit{sw},\mathit{snic}} \le 21.3$ Gbps even without any filtering (i.e., when $p = q = 0$ and therefore $r = 1$). Thus, we expect no packet loss on the $B_{\mathit{sw},\mathit{snic}} = 100$ Gbps link.

\subsubsection{Reduction of Load on \snic by Packet Filters:}
\label{sec:filter-rate-eqs}
The discussion above eliminates concerns about bandwidth limitations on the \snic link. However, we must still ensure that the \snic's packet-processing capacity ($\mu_{\mathit{snic}}$) is not exceeded. Based on Eq.~\eqref{eq:r}--\eqref{eq:lsnic}, satisfying $\lambda_{\mathit{snic}} \le 53$ MPPS requires that $r \le 0.42$.

Figure~\ref{fig:snic-load} illustrates the \snic load and the reduction achieved because of the hit rate on the MST and OFF tables measured in our simulator; the dashed horizontal line indicates the maximum processing capacity of the \snic, and we seek to ensure the delivered load ($\lambda_{\mathit{snic}}$) is below this.
In the simulator, we obtain the hit ratios $p$ and $q$ for the MST and OFF under a large-scale traffic scenario with millions of concurrent flows that would be expected with Tbps-level traffic. We simulate benign flows with an average of 53 packets each, while malicious flows consist of single-packet SYN scans, with each of 10,000 malicious hosts scanning 600 targets. We model flow dynamics using a birth–death process: 7 million benign flows are maintained concurrently, with a new flow starting when another ends. Likewise, a new malicious host begins scanning when another completes. Within a one-second window, the traffic contains 5.6 million unique benign flows and 2.5 million unique malicious flows. We allocate 40 KB of TCAM to the MST and 4 MB of SRAM to the OFF table, matching our testbed experiment.

Figure~\ref{fig:snic-load} shows the MST and OFF reduce the \snic load sufficiently:
With the MST alone (representing a fixed $q=0$, leftmost group of bars), the malicious filter is able to redirect about 35 MPPS (achieving $p=0.99$), but this still leaves 90 MPPS to be processed.
With only the OFF ($p=0$, middle group), the benign filter can provide even greater savings since most packets are from normal flows, but the residual rate to be processed by \snic ($\lambda_{\mathit{snic}}$) would exceed its capacity.
However, \name by deploying the MST followed by the OFF (rightmost group) effectively redirects both malicious and benign flows before they reach the \snic. Our simulation shows that $p=0.99$ and $q=0.61$.  Thus, we ensure that there is no loss of packets sent to \snic processing in our design for the expected filter hit rates.

\subsubsection{Malicious Source Update over the Control Plane:}
We next validate our assumption that updating the switch TCAM with the identity of a malicious source is acceptable even via the relatively slower control plane.
Figure~\ref{fig:update} presents the update frequency for MST ($\lambda_{cpu}$) and the update speed via the control plane ($\mu_{cpu}$).
The update frequency never exceeds the maximum control plane update speed, even when the traffic volume scales up to 1.5 Tbps.
Further, we also plot the required update frequency for OFF, which significantly exceeds the control plane speed, but is below a single port's bandwidth.
This supports our design choice to use data plane to update the OFF table, while control plane for MST.
\begin{figure}[t]
    \centering
    \begin{subfigure}[t]{0.48\linewidth}
         \includegraphics[width=\linewidth]{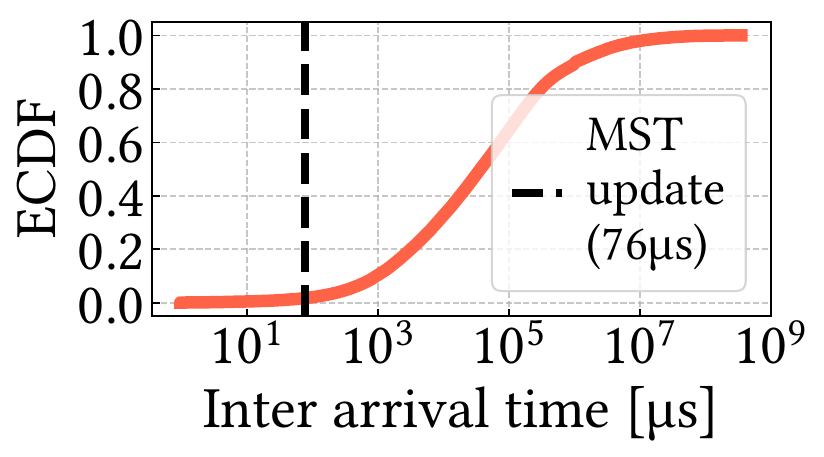}
        \caption{Scan (MST update)}
        \label{fig:mst-vs-scanner}
    \end{subfigure}
    \hfil
    \begin{subfigure}[t]{0.48\linewidth}
         \includegraphics[width=\linewidth]{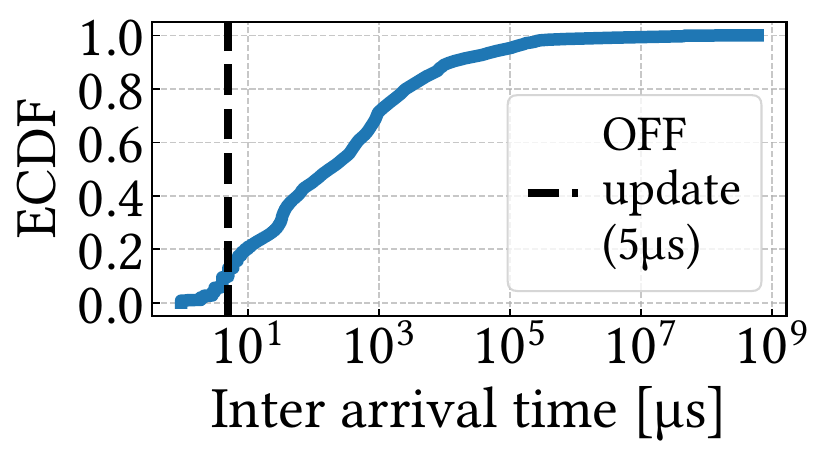}
        \caption{Benign flow (OFF update)}
        \label{fig:off-vs-benignflow}
    \end{subfigure}
    \caption{MST/OFF update latency vs. packet inter-arrival time (IAT) for scan and benign flow in the MAWI Dec. trace. Both update latencies are low enough for the switch to react before the next packet arrives after \snic detection.}
    \label{fig:update-latency-measurement}
\end{figure}

\subsubsection{MST/OFF Update Latency}
We next validate whether MST and OFF updates are fast enough for the switch to react before the next packet arrives after detection at the \snic. For this validation, we measure the update latency of MST and OFF on our testbed and analyze packet inter-arrival times (IATs) for each scan source and each benign flow in the MAWI Dec. trace. 

Figures~\ref{fig:mst-vs-scanner} and~\ref{fig:off-vs-benignflow} show the ECDFs of IATs for the scan packets and benign flow packets, respectively. The MST and OFF updates take an average of $76\,\mu\mathrm{s}$ and $5\,\mu\mathrm{s}$, respectively. While MST updates traverse the control plane and are thus slower than OFF updates, 98.3\% of scan packet IATs are more than the MST update latency. Likewise, 90.5\% of benign flow packet IATs exceed the OFF update latency. These results indicate that MST and OFF updates are sufficiently fast relative to packet arrivals in real traffic, enabling \name to react promptly after \snic detection.
\section{Evaluation}
\label{sec:eval}

In addition to our hardware implementation in \S~\ref{sec:implementation}, we develop the full \name system in two simulators: one accepting packet traces and one for synthetic traffic. The trace-based simulator evaluates ML effectiveness using Zeek-labeled ground truth, while the synthetic one stress-tests the system at higher traffic rates using a perfect classifier. 

We evaluate using the CAIDA and MAWI traces, as well as the CIC-IDS2017~\cite{CICIDS_2017} traces which contain artificially generated Port Scan and Slowloris attacks (allowing for a perfect ground truth). These traces can be used in simulation, but we cannot replay them at sufficient speed to saturate our hardware testbed; for that purpose we use synthetically generated mixes of both benign and attack traffic.

\begin{figure}[t]
    \centering
        \captionsetup[subfigure]{skip=-4pt}
    \begin{subfigure}[b]{0.50\linewidth}
        \includegraphics[trim=0.3cm 1cm 0.95cm 0.8cm, clip, width=\linewidth]{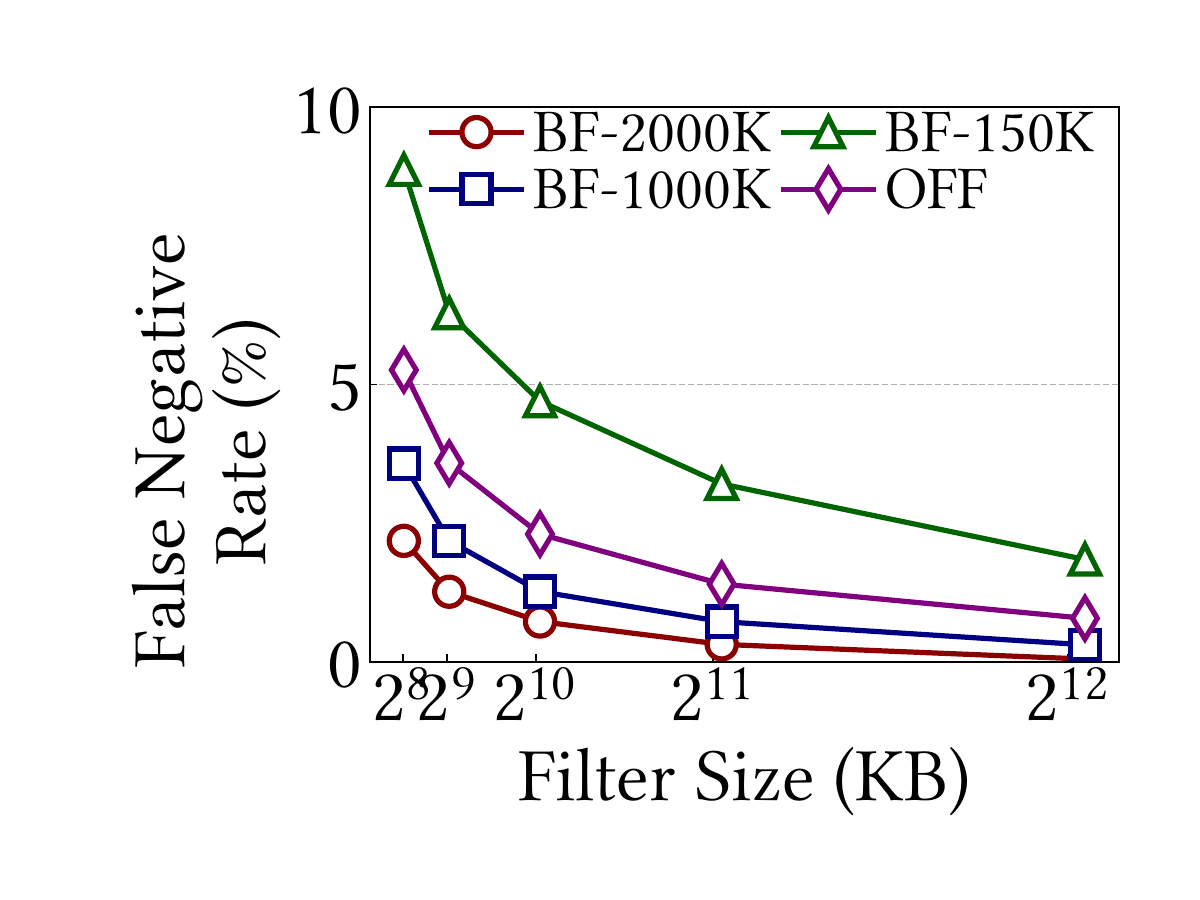}
         \caption{MAWI Trace}
        \label{fig:off-bf-fpr-mawi}
    \end{subfigure}
    \hfil
    \begin{subfigure}[b]{0.50\linewidth}
        \centering
        \includegraphics[trim=0.3cm 1cm 0.95cm 0.8cm, clip, width=\linewidth]{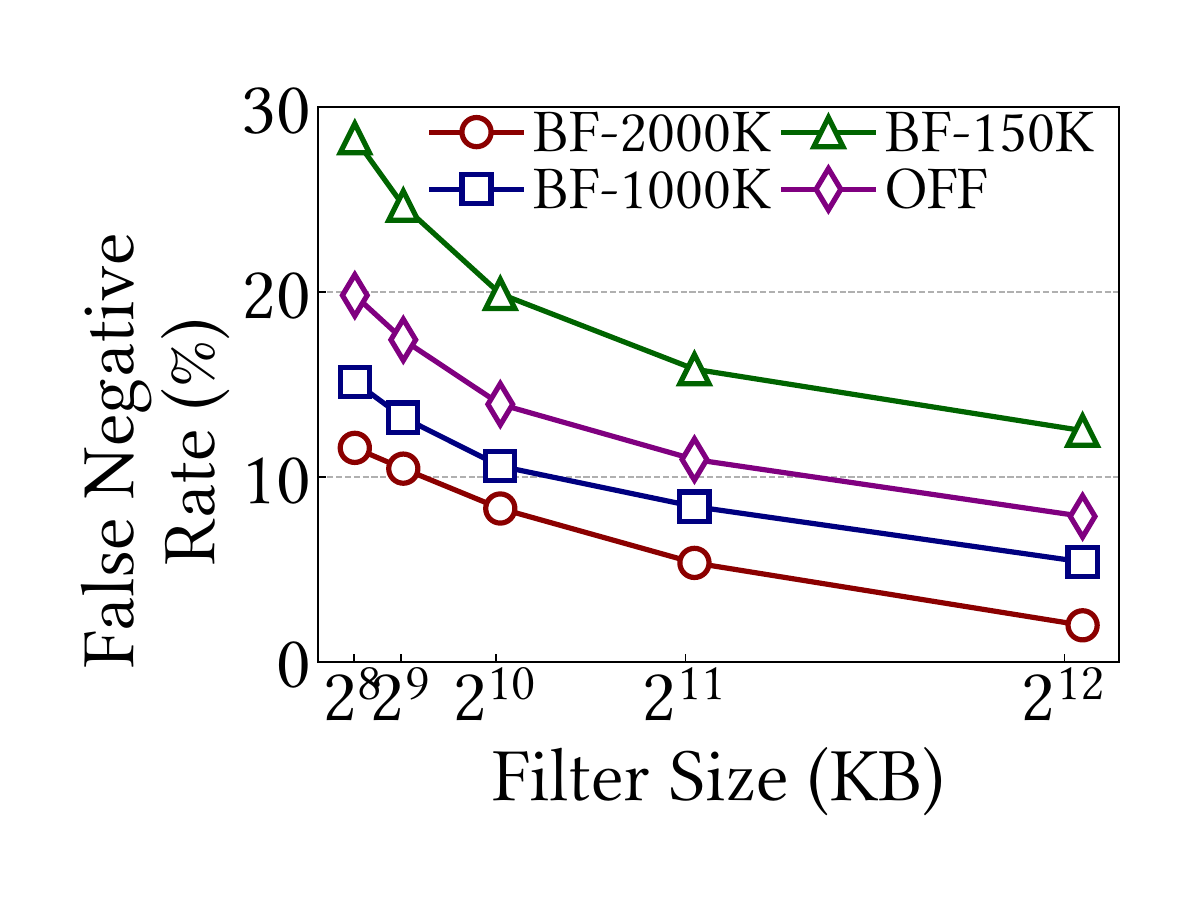}
        \caption{CAIDA Trace} 
        \label{fig:off-bf-fnr-mawi}
    \end{subfigure}
    \caption{FNR: Aging Bloom Filter vs. OFF}
    \label{fig:off-bf-fnr}
\end{figure}

\subsection{Filter Performance}

We first evaluate the False Positive and False Negative Rates for our OFF data structure using our simulation platform when running real Internet traces. 
We compare against an Aging Bloom Filter (BF) data structure modeled after~\cite{5066970}, where a primary Bloom filter is periodically swapped with a secondary, allowing active flows (hits in the filter) to be propagated while entries that are not hit are evicted. While this allows cells to be naturally evicted based on recency, it halves the available memory. 
This means that while the Bloom filter could potentially store more entries (reducing false negatives), it is harder to detect collisions (increasing false positives).
The total memory for both filters is the same (4 MB), but we vary the maximum number of entries that can be stored in the Bloom Filter before the swap occurs, e.g., `BF-150K' and `BF-2000k' are Bloom Filters that will cycle their tables after 150 thousand and 2 million insertions, respectively. The OFF table can store a total of 2 million entries with this memory size and doesn't need to reset.

To stress test the filters, we speed up the MAWI and CAIDA traces by breaking them into chunks which are then interleaved; this leads to a higher flow concurrency and bit rate while keeping flow completion times the same as the original trace. Our combined MAWI trace has 321 Gbps of traffic while CAIDA has 395.6 Gbps.

\begin{center}
  \centering
  \begin{small}
  \vspace{-2mm}
  \begin{tabular}{@{}lllll@{}}
    \toprule
    & OFF & BF-2000K  & BF-1000K  & BF-150K\\ \midrule
    MAWI Trace & 0.0014\% & 2.65\% & 1.17\% & 0.22\%\\
    CAIDA Trace & 0.0008\% & 1.21\% & 0.36\% & 0.005\%\\ \bottomrule
  \end{tabular}
  \vspace{-1mm}
  \end{small}
\end{center}
The table above shows the False Positive Rates (FPR, the number of unknown packets mistaken as benign due to collisions), note the log scale. OFF provides at least an order of magnitude improvement in FPR, reaching a rate of only 0.0008\% of packets. We separately find that the FPR values are stable with table size (as expected by our FPR equation), only rising to 0.0014\% when the memory size is 128KB. 

In Figure~\ref{fig:off-bf-fnr} we evaluate the impact of filter size on False Negative Rate (FNR, number of benign flows mistaken as unknown by the filter due to evictions). Here we see the benefit of the Bloom Filter being able reduce collisions by only using 2 independent bits to store each entry. However, the $<$ 5\% improvement in False Negatives that the Bloom Filter provides is less important than OFF's 150x to 1890x improvement in FPR since in our problem domain false negatives simply mean more work for the \snic, while false positives indicate potential security compromises.

\subsection{Real Trace Filtering Effectiveness}

We use our trace-driven simulator to measure the percent of packets analyzed at each component in \name. Since the traces are at a lower rate, we downscale the OFF size by a factor of 1024.
We get the following breakdown:

\begin{center}
  \centering
  \label{tab:immunity_breakdown}
  \begin{small}
    \vspace{-2mm} 
  \begin{tabular}{@{}ccccc@{}}
    \toprule
    & Hit on MST  & Hit on OFF  & To SmartNIC \\\midrule
    CIC Port Scan & \textcolor{OliveGreen}{9.09\%} + \textcolor{red}{0\%} & \textcolor{OliveGreen}{73.45\%} + \textcolor{red}{1.03\%} & 16.43\%\\
    CIC Slowloris & \textcolor{OliveGreen}{0.61\%} + \textcolor{red}{0\%} & \textcolor{OliveGreen}{98.01\%} + \textcolor{red}{0\%} & 1.38\%\\
    MAWI 12/16 & \textcolor{OliveGreen}{26.5\%} + \textcolor{red}{3.3\%}   & \textcolor{OliveGreen}{59.8\%} + \textcolor{red}{0.08\%} & 10.3\% \\
    MAWI 11/15 & \textcolor{OliveGreen}{23.8\%} + \textcolor{red}{2.3\%} & \textcolor{OliveGreen}{63.2\%} + \textcolor{red}{0.06\%} & 10.6\% \\
    CAIDA & \textcolor{OliveGreen}{14.0\%} + \textcolor{red}{0.01\%} & \textcolor{OliveGreen}{39.16\%} + \textcolor{red}{0.03\%} & 46.8\% \\
    \bottomrule
  \end{tabular}
  \end{small}
\vspace{-1mm} 
\end{center}

For traffic sent to the Scrubber (after a hit on the MST), we separate out the volume of correctly handled suspicious traffic (i.e., malicious packets, incomplete and non-benign flow packets, and packets from malicious IPs; labeled in green) and the amount of missclassified traffic (i.e., traffic from benign IPs classified by our 'online' Zeek rule as bad; labeled in red). For example, MAWI 12/16 shows the MST correctly filtering 26.5\% of the traffic entering the switch as being suspicious, while 3.3\% of the switch's traffic is misidentified packets from benign IPs, typically because the source IP had enough incomplete flows in our online approximation that it was marked as a potential scanner by rules on the SmartNIC. The CIC traces indicate perfect accuracy for \name since the attacks are relatively light and easy to differentiate. For all traces we see < 4\% error rates, which we believe is acceptable since this only represents good traffic that will see unnecessary monitoring.

The majority of traffic (except in CAIDA, see below) is sent directly to the destination after hitting in the \off, illustrating \name's potential to rapidly forward good traffic without wasting monitoring cycles. However, a small fraction of suspicious traffic (0.03\% to 1.03\%) is sent out directly because of collisions in the \off structure or missclassifications by the \snic ML. We believe CIC Port Scan does poorly due to ML misclassifications marking some attacks as benign; this could be improved by training a classifier specific to the trace or tuning the IDS rule parameters.

Finally, the SmartNIC must process at most 16.43\% of traffic for MAWI and CIC. CIC Slowloris has very little NIC traffic since most of the trace is long running benign flows with only a few, easy to detect attacks. We find that about half of the packets reaching the NIC are from benign flows (for which we must observe the first few packets to identify the completed handshake) and half are incomplete flows or attacks that are later added to the MST. This illustrates \name's potential to substantially reduce the amount of traffic that needs deep analysis on the NIC.

CAIDA has fewer packets sent directly via the \off (38\%) and more needing to go through the SmartNIC (46.8\%) because the trace has an unusually high volume of Incomplete (but not deemed as malicious) flows. We find that about 8\% of the packets reaching the \snic are from benign flows, a similar ratio as in MAWI/CIC-IDS, but about 23\% are from incomplete flows not deemed malicious since they were below our thresholds. 
However, even in this workload where more than half of the traffic is suspicious packets, we estimate that \name would be able to meet close to line rate. Our analysis from \S\ref{sec:filter-rate-eqs} shows we could process 1.5 Tbps if less than 42\% of traffic needed to visit the SmartNIC; thus this workload, sending 46.8\% to the NIC, would only lower our maximum supportable rate to 1.3 Tbps.

\begin{figure}[t]
  \centering
  \begin{minipage}{0.25\textwidth}
    \centering
    \includegraphics[trim=0.1cm 1cm 0.1cm 0.8cm, clip, width=\linewidth]{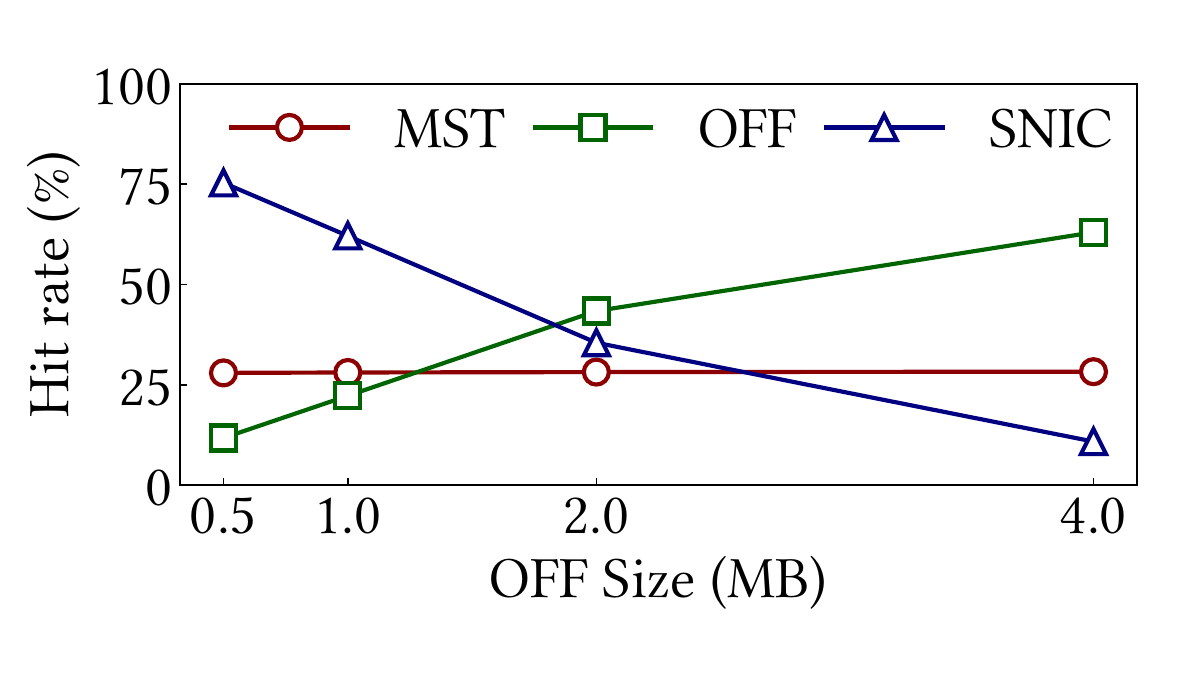}
    \caption{ OFF Size vs Hit Rate}
    \label{fig:offsize_vs_rate}
  \end{minipage}
  \begin{minipage}{0.22\textwidth}
    \centering
    \includegraphics[width=\linewidth]{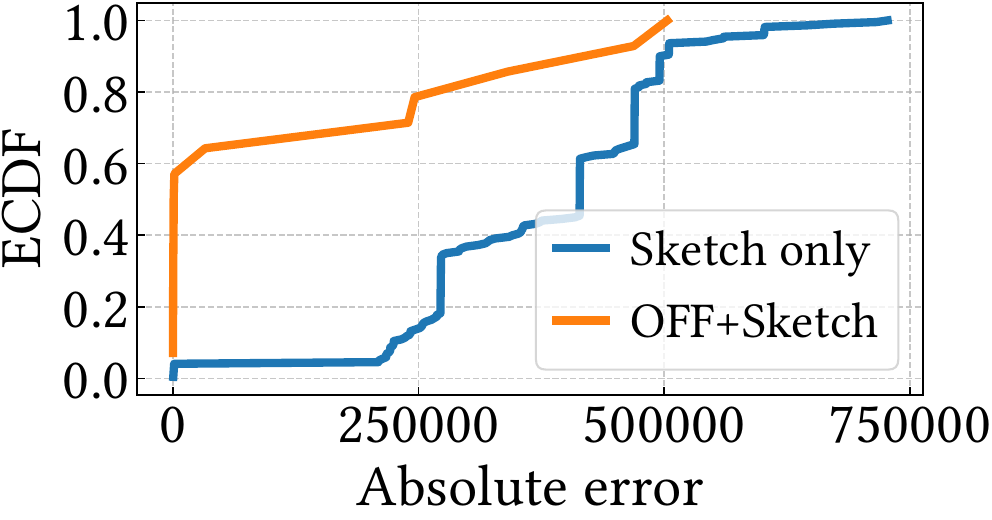}
    \caption{Sketch Error}
    \label{fig:sketchcdf}
  \end{minipage}
\end{figure}

To validate that this behavior is representative with larger traffic rates and filter sizes, we used our second simulator to measure the impact of OFF size on the traffic breakdown.
We generate 1.5 Tbps of traffic with a flow mix similar to MAWI--3.2 million benign flows per second and 7.1 million malicious flows per second--and measure the \% of packets hit in MST, OFF, or sent to SmartNIC.
As shown in Figure~\ref{fig:offsize_vs_rate}, with the full size OFF and full rate traffic, the \snic needs to handle only 10.9\% of the packet load, very similar to the amount observed in our trace-driven experiment at lower rate/size.
When the OFF size is reduced to only 2 MB, then 35.5\% of packets must be delivered to the NIC (as batched headers) since the benign filter is less effective.

\begin{figure*}[t]
    \begin{minipage}[b]{0.63\textwidth}
        \begin{footnotesize}
        \begin{tabular}{lccc|ccc}
        \toprule
        & \multicolumn{3}{c|}{\textbf{CAIDA (Precision \textbar{} Recall)}} & \multicolumn{3}{c}{\textbf{MAWI (Precision \textbar{} Recall)}} \\
        \textbf{Attacks} & \textbf{NetBeacon} & \textbf{NB+IDS} & \textbf{immUNITY} 
                      & \textbf{NetBeacon} & \textbf{NB+IDS} & \textbf{immUNITY} \\
        \midrule
        Scans & 0.04 \textbar{} 0.54 & 0.45 \textbar{} 0.43 & 0.95 \textbar{} 0.99 & 0.28 \textbar{} 0.59 & 0.93 \textbar{} 0.53 & 0.99 \textbar{} 0.99 \\
        Dist Syn & 0.00 \textbar{} 0.00 & 0.00 \textbar{} 0.00 & 0.99 \textbar{} 0.95 & 0.00 \textbar{} 0.00 & 0.00 \textbar{} 0.00 & 0.98 \textbar{} 0.93 \\
        Slowloris & 0.003 \textbar{} 1.00 & 0.57 \textbar{} 0.80 & 1.00 \textbar{} 1.00 & 0.001 \textbar{} 1.00 & 0.15 \textbar{} 0.80 & 1.00 \textbar{} 1.00 \\
        SSH Brute & 0.77 \textbar{} 1.0 & 0.88 \textbar{} 0.70 & 1.00 \textbar{} 1.00 & 0.19 \textbar{} 1.00 & 0.88 \textbar{} 0.70 & 1.00 \textbar{} 1.00 \\
        FTP Brute & 0.03 \textbar{} 1.00 & 0.91 \textbar{} 1.00 & 1.00 \textbar{} 1.00 & 0.02 \textbar{} 1.00 & 0.91 \textbar{} 1.00 & 1.00 \textbar{} 1.00 \\
        \bottomrule
        \end{tabular}
        \end{footnotesize}
    \end{minipage}
    \hfill
    \begin{minipage}[b]{0.3\textwidth}
        {\includegraphics[height=1.2in]{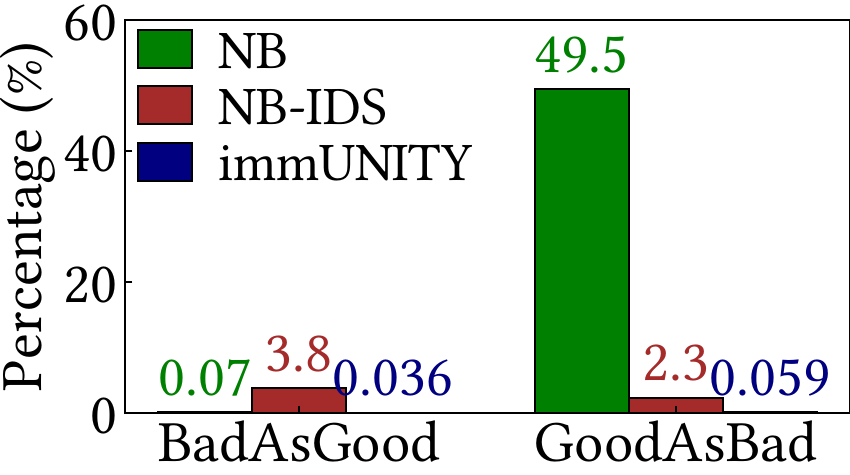}}
    \end{minipage}
    \hfill
    \hfill
    \caption{{\em Table:} Accuracy of malicious IP detection by NetBeacon, NetBeacon with \name's IDS rules (NB+IDS) or \name. {\em Figure:} \name significantly reduces the number of malicious IPs treated as good and vice versa in CAIDA. }
    \label{tab:caida-mawi-comparison}  
    \vspace{-4mm}
\end{figure*}

\subsection{Attack Detection Accuracy Comparison}
\label{sec:MST-Detection}

\head{NetBeacon:} We compare the accuracy of \name in detecting malicious sources against NetBeacon~\cite{zhou_efficient_2023}, a system that deploys both a stateless packet model and a stateful flow model on a P4 switch.
NetBeacon does not require a \snic like our system, but it must use simpler decision tree ML models to fit within the switch's constraints. Further, its flow table is limited by the switch's memory, and when this becomes full more packets will be processed by the less accurate packet model. NetBeacon only attempts to report classifications for individual \textit{packets or flows}, so we also build an enhanced version of NetBeacon, integrating \name's IDS rules with it to create a system that can identify malicious \textit{sources} using the same attack thresholds as \name. 

Figure~\ref{tab:caida-mawi-comparison} shows the accuracy in terms of malicious source IPs detected by NetBeacon (which treats the source IP of any packet classified as an attack as a malicious IP), NB+IDS (which uses NetBeacon's packet classification system, enhanced with our IDS thresholds to identify malicious sources), and \name. We monitor each of the attack types in the CAIDA and MAWI traces with our synthetic  attacks added. 

NetBeacon has moderate Recall (it finds 54-59\% of scans, and all of the Slowloris, SSH, and FTP attacks), but it  misclassifies many benign flows as malicious, leading to extremely low precision scores ($4-28\%$ for scans, and as low as 0.001\% for slowloris). This emphasizes the importance of statefully aggregating multiple flow decisions to identify malicious sources as done by our IDS rules. Without this, NetBeacon incorrectly marks 49.5\% of benign IPs as malicious in CAIDA as shown at right in Figure~\ref{tab:caida-mawi-comparison}.
By leveraging our IDS rules, NetBeacon+IDS can be substantially more precise, but it still misses some types of attacks entirely. In terms of aggregate detection, NetBeacon+IDS still misses 3.8\% of attackers and 
mislabels  2.3\% of benign IPs as malicious in CAIDA. 

In contrast, \name maintains precision and recall above 93\%
since using the \snic provides sufficient memory for flow table state and the necessary computation for our more complex ML model. In CAIDA, \name misidentifies only 0.036\% of attackers, and only marks 0.06\% of the benign IPs as needing analysis by the scrubber (improvements of 105x and 38x respectively). The high accuracy demonstrates that our system can effectively identify attackers, even those who launch low-volume or slow attacks.

\head{Sonata:} We also compared against the Sonata~\cite{gupta_sonata_2018} switch-based query system for detecting port scans and slowloris attacks. While Sonata achieves good accuracy on traffic from its own generation scripts, it is not as good on real traces. On a subset of the MAWI trace, Sonata achieves only an F-1 score of 0.36 vs \name's 0.97. 

\head{Heavy Hitter Detection:} Finally, we evaluate the impact of deploying a heavy hitter sketch as part of \name's pipeline. We first evaluate the precision of the sketch in identifying heavy hitter flows (configured as flows with > 200K packets). We find that using the sketch by itself, results in poor precision--the sketch detects 220 flows as heavy, but in reality only 9 are correct, leading to a precision of 4.1\%. In contrast, when we deploy the sketch after \name's benign flow filter, the precision is improved to 64.3\%, reducing the number of misidentifed heavy flows by 97.6\%. This shows the value of filtering out the massive number of scan packets which add noise to the sketch resulting in many false positives. Figure~\ref{fig:sketchcdf} shows the ECDF of the absolute error in counts per flow for the two approaches, illustrating how using the \off greatly improves accuracy.

\subsection{Testbed Experimental Results}
\label{sec:testbed-experiment}

We construct the testbed system from Fig.~\ref{fig:prototype} consisting of a single switch 100 Gbps port, a single \snic core, and MST/OFF reduced to 1/16 of their full size. The switch is connected to \snic by a single 100 Gbps link. This configuration represents 1/16 of a full system deployment (having 16 switch ports and 16 \snic cores). We show the per-core processing behavior of the full system. We inject 100 Gbps traffic with 100K concurrent flows corresponding to a 1/16-scaled load of the 1.6 Tbps traffic in the full deployment. The traffic consists of a mix of benign flow packets and scan packets in a 10:1 ratio, using 1 KB packets.

\head{Results:}
Under this configuration, \name achieves 96 Gbps (11.7 Mpps) of end-to-end throughput without packet loss, while 85\% and 8\% of the total traffic hit the OFF and MST, respectively. As a result, the switch redirects only 0.14 Gbps to the \snic, which delivers packet information at a rate of 0.81 MPPS (a batch of three packets at 0.27 MPPS). Even when scaling to 16 \snic cores, the total redirected traffic is only 2.24 Gbps, which is below the 100 Gbps switch–\snic link capacity. These results demonstrate that \name can run at line rate without over-saturating the \snic’s bandwidth and packet processing capacity.

\head{Scalability to \snic Multi-Cores:}
Our current implementation is single threaded, but further development effort would allow it to scale up to a full 16-core deployment. The flow table can be fully sharded on a per-flow basis using RSS, allowing each \snic core to maintain its own state exclusively. Second, the ML model is shared across cores in the \snic but accessed in a read-only manner, eliminating the need for mutual exclusion such as locking. \name's throughput should scale nearly linearly with the number of \snic cores, making Tbps-class traffic handling feasible.
\section{Conclusions}

To detect low volume and slow attacks, we flip the design philosophy of typical IDS/IPS designs and
efficiently identify and fast-path \textit{benign} traffic directly in a switch dataplane. This drastically reduces the volume of traffic requiring deep, stateful analysis on a \snic. Combining sketch-based volumetric monitoring for high-rate attacks with fine-grained, flow-level ML classification and cross-flow rule-based detection of low volume and slow attacks, we achieve comprehensive coverage across diverse threat types. 

Our design shows the importance of making data structures that fit within the constraints of typical programmable switches and the optimization of communication between the switch and the SmartNIC. We can efficiently track millions of flows with minimal memory footprint and low false positive/negative rates.
We achieve update latencies of 5 and 75 microseconds via the data and control planes respectively, ensuring real-time responsiveness even under high load.

We show, using trace-driven simulation and on our hardware testbed, that \name can accurately identify 99\% of scan attacks and can scale to support 1.3-1.5 Tbps of traffic, without overloading the \snic. This lets \name match the performance of switch-only solutions like NetBeacon, while lowering the percent of missclassified attackers from 3.8\% to 0.036\%.
\bibliographystyle{ACM-Reference-Format}
\bibliography{reference,gw-zotero}

\appendix
\clearpage
\appendix

\section{Appendix}
\label{sec:appendix}

\subsection{Trace Analysis}

The MAWI trace below is used for testing in some of our experiments, while a portion of MAWI Dec. (Table~\ref{tab:mawi_1216}) is used for training.
\begin{table}[H]
\centering
\caption{MAWI Trace (Nov. 15, 2024) [14:00-14:15].}
\begin{small}
\begin{tabular}{ l ccc } 
  \toprule
 Field & IPs & Flows & Packets \\
  \midrule
  Total & 587,160 & 14,937,679 & 69,971,130 \\
  Scan \% & 2.5\% & 69.1\% & 15.5\% \\
  Benign & 41,128 & 547,994 & 53,509,590 \\
  Incomplete & 419,069 & 3,965,706 & 4,741,629 \\
  Non-Benign & 6,946 & 104,984 & 1,005,331 \\
  Scan & 492,997 & 10,318,995 & 10,827,444 \\
  Distributed Syn & 1,925 & 25,067 & 128,810 \\
  \bottomrule
\end{tabular}
\label{appendix:mawi_1115}
\end{small}
\end{table}

\subsection{Online Scan Algorithm}
\label{sec:online-detection}
\begin{algorithm}[H]\
\caption{immUNITY Online Scan Detection}
\label{alg:immunity-scan}
\begin{small}
\begin{algorithmic}[1]
    \State \textbf{Input: } pkt
    \State \textbf{Output: } a list of scanners
    \State \textbf{Initialize} 
    \If{$scanner \in known\_scanners$} 
        \State \Return {$known\_scanners$}
    \EndIf
    \State let $scan\_info$ be a list of possible $scanners$
    \State let $attempt\_key$ be ($victim, scanned\_port$)
    \If{first packet to $attempt\_key$ and $syn\_flag$}
        \State Start tracking the scanning $attempt$
        \State Check if scanning thresholds are exceeded (Allows 1 sec delay per flow)
            \If {Threshold exceeded}
                \State $known\_scanners$.add($scanner$)
            \EndIf
        \State \Return {$known\_scanners$}
    \EndIf
    \If{$response$ and not $rst\_flag$} 
    \State Remove $scanner$ from $scan\_info$ \EndIf
    \State Check if scanning thresholds are exceeded (Allows 1 sec delay per flow)
    \If {Threshold exceeded}
        \State $known\_scanners$.add{$scanner$}
    \EndIf
    \State \Return {$known\_scanners$}
\end{algorithmic}
\end{small}
\end{algorithm}

This pseudocode is for our algorithm to run an online scan detection based on a Zeek rule. The modification we needed to add for our online version includes a small delay (1 second) for a flow before concluding that it is an incomplete flow without completing the connection setup. We also run a check every 15 trace seconds to timeout any flows and evaluate all threshold conditions. We develop similar algorithms for each attack and deploy to the \snic. For a Distributed SYN attack, we modified the detection from sampling based detection to analyzing every packet that the \snic sees. Similar to the case with scans, a small delay (1 second) is applied for timing out an incomplete flow. Slowloris and SSH Bruteforce follow the same thresholds as Zeek but the attacker IP is recorded in the MST immediately after the thresholds are met.

\subsection{ML Model Selected Features}
\label{sec:ml-snic}
We base our \snic ML model on Leo~\cite{jafri24:leo}. Leo applies recursive feature elimination to select the most effective features for classification. To balance accuracy and processing cost, we configure the model to use ten features and a decision-tree depth of eight.
For training, we use the MAWI December trace. 
Table~\ref{tab:ml-features} lists all the features we used to train our final ML model. The feature set was chosen based on recursive feature elimination mechanism, which is used in Leo~\cite{jafri24:leo} from flow-level features defined in CICFlowMeter~\cite{CICFlowMeter}. 

\begin{table}[t]
\centering
\caption{Selected Features for ML Model}
\begin{tabular}{ |l|  } 
  \hline
  Description  \\
  \hline
    Destination port \\ 
    Flow duration \\
    Total packet length in the forward direction \\
    Total packet length in the backward direction \\
    Minimum packet length in the forward direction \\
    Minimum flow inter-arrival time \\
    Maximum packet length \\
    TCP RST flag count \\
    TCP PSH flag count \\
    TCP ACK flag count \\
  \hline
\end{tabular}
\label{tab:ml-features}
\end{table}

\section{Ethics}
We conducted this study in accordance with the ethical guidelines of all involved institutions as well as those specified by the dataset providers. Specifically, the analysis relies on publicly available packet traces from the MAWI project and CAIDA, both of which have been anonymized prior to release and contain no personally identifiable information. At no point did we attempt to infer or reconstruct any personal information from these datasets. In addition, we evaluated the proposed system using these anonymized traces and synthetic packet traces. For these reasons, we believe that this work does not raise any ethical concerns.

\end{document}